# Crystallography of periodic nanotextures in a strained Mott insulator

*Benjamin Z. Gregory[1,2], Yorick A. Birkhölzer[1], Noah Schnitzer[1], Ziming Shao[1], Jeff Hodgson[1], Suchismita Sarker[3], Jacob P. Ruff[3], Berit H. Goodge[4], David A. Muller[5,6], Kyle M. Shen[2,6], Darrell G. Schlom[1,6,7], Andrej Singer[1,*]*

[1]*Department of Materials Science and Engineering, Cornell University; Ithaca, NY 14853, USA*
[2]*Department of Physics, Cornell University; Ithaca, NY 14853, USA*
[3]*Cornell High Energy Synchrotron Source, Cornell University; Ithaca, NY 14853*
[4]*Max Planck Institute for Chemical Physics of Solids; 01187 Dresden, Germany*
[5] *School of Applied and Engineering Physics, Cornell University; Ithaca, NY 14853, USA*
[6]*Kavli Institute at Cornell for Nanoscale Science, Cornell University; Ithaca, NY 14853, USA*
[7]*Leibniz-Institut für Kristallzüchtung; Max-Born-Straße 2, Berlin 12489, Germany*
[*]*asinger@cornell.edu*

## Abstract

Here we investigate stripes of alternating structural phases spontaneously forming in epitaxially strained $Ca_2RuO_4$ thin films below the metal-insulator transition. Using large-volume X-ray reciprocal-space mapping, we show that satellite-pattern intensities across 24 symmetry-inequivalent Bragg reflections collapse onto a single parameter-free curve. The collapse identifies a coherent martensitic laminate of few-nm wide domains separated by $\{012\}$ interfaces, with displacements along $\langle 01\bar{2}\rangle$. Satellite-extinction analysis demonstrates that both coexisting phases retain the bulk orthorhombic space group despite the pseudocubic $LaAlO_3$ substrate, biaxial epitaxial strain, and intrinsic strain at the interfaces. Classical invariant-plane-strain crystallography thus governs the nanoscale domain geometry of a Mott insulator with intertwined magnetic, electronic, and lattice order.

## Main Text

### *Introduction*

Almost a century ago, Landau and Lifshitz showed that ferromagnetic domains arise from competition between demagnetization energy and domain-wall energy, with the domain width scaling as the square root of crystal thickness [1]. Kittel extended the analysis to thin films and small particles [2]. The same competition explains ferroelectric domains, with polarization in place of the magnetic energy [3], and ferroelastic domains in epitaxial thin films, with elastic strain energy across coherent twin interfaces [4]. A universal scaling law of domain width and film thickness was shown to hold for ferroics [5]. Strongly correlated oxides display similar domain morphologies forming at metal-insulator transitions [6–9], such as martensitic-like twin textures in charge ordered manganites [10]. Unlike in ferroics, however, the coexisting domains are not symmetry-related variants of one phase. Instead, they are distinct phases with different lattice, electronic, and magnetic order. One line of theory describes the emergence of domain patterns through coupling between unit-cell lattice distortions and long-range elastic strain [11], while another attributes the domain morphologies to electronic phase competition and quenched disorder [12]. Yet the three-dimensional structure of the coexisting phases, their interface crystallography, and their elastic compatibility have remained largely inaccessible to direct structural probes, obscuring the microscopic origin of the emergent domain patterns.

The Ruddlesden–Popper compound $Ca_2RuO_4$ is a canonical Mott insulator: above 357 K it is a paramagnetic metal; below 357 K it is an antiferromagnetic insulator. During the temperature–induced transformation, the lattice deforms from long *c*-axis (L-*Pbca*) high-temperature phase (Fig.1a) to the short *c*-axis (S-*Pbca*) low-temperature phase. During the L to S transformation, the orthorhombic lattice parameters change by a few percent (Table S1), the orthorhombic *Pbca* space group is preserved [13–15], and the octahedra rotate and flatten [16,17]. The strong coupling between electronic order and structural distortions makes the electronic order tunable via hydrostatic pressure [18,19] (with phase coexistence at around 1 GPa [19]) and biaxial epitaxial strain [20–22]. Consistently, a compressive strain caused by epitaxial growth on $LaAlO_3$ of -1.5% (Table S1) lowers the insulator-to-metal transition temperature of $Ca_2RuO_4$ thin films by more than 100 K [9,20] (Fig. 1a). Recent X-ray nanoimaging and electron microscopy revealed that these strained $Ca_2RuO_4$ films, homogeneous in the high-temperature phase, spontaneously form periodic striped nanodomains upon cooling, with alternating layers resembling S-*Pbca* and L-*Pbca* [9,23,24] (Fig.1a, inset). Striped nanodomains also form at the surface of bulk $Ca_2RuO_4$ crystals during a current-induced transition [25,26]. The recurrence of striped nanodomains across temperature-induced transitions in strained films and current-induced transitions in bulk suggests that nanodomain formation is a general response of this system.

Three techniques have probed striped nanodomains in $Ca_2RuO_4$, each limited to a partial view of the structure. Scanning near-field optical microscopy [25] resolves variation of optical properties across the domains but not the underlying atomic structure. Single-Bragg-peak X-ray diffraction [9] couples to only one projection of the displacement field. Transmission electron microscopy [9,23] returns a two-dimensional projection of a three-dimensional structure. Here we map a continuous $(5\ \text{Å}^{-1})^3$ volume of reciprocal space containing more than 100 reciprocal lattice points from a biaxially strained $Ca_2RuO_4$ film, where satellite peaks around reciprocal lattice points capture domain morphology. While large 3D reciprocal space mapping is becoming an established technique to study thin films [27], the analysis of these data remains challenging [28] in part because satellites from different coexisting interfaces overlap and interfere. The analysis often involves combined phase field modelling of real space and diffraction simulations [29,30]. We show that a closed-form scattering model built on invariant-plane-strain martensitic theory [31] reproduces the full reciprocal-space dataset without fitted parameters. We further use the satellite intensities to show that the orthorhombic symmetry persists inside both phases of the striped domains despite the pseudocubic substrate, biaxial epitaxial strain, and strain due to the phase transformation.

## *Invariant plane strain framework*

The satellite patterns around multiple Bragg peaks collectively encode the relative displacements inside the coexisting structures and the structure of the interfaces. We describe the framework that predicts their shapes using concepts from the theory of martensitic transformations. In this framework, coherent interfaces between coexisting structures accommodate the transformation strain through an invariant plane strain (IPS) deformation [31–34]. A specific crystallographic plane (invariant plane with normal $\mathbf{n}$) remains undistorted and unrotated, while the lattices on either side undergo displacement along vector $\mathbf{l}$ (Fig. 1b). In reciprocal space, the two structures generate two distinct reciprocal lattices, with the shift, $\boldsymbol{\Delta}_{\mathbf{G}}$, between corresponding Bragg peaks given by [31]

$$(1) \quad \boldsymbol{\Delta}_{\mathbf{G}} = -\varepsilon \boldsymbol{n} \cdot (\boldsymbol{G} \cdot \boldsymbol{l}) / \big(1 + \varepsilon(\boldsymbol{n} \cdot \boldsymbol{l})\big),$$

where $\varepsilon$ is strain, and $\mathbf{G}$ is the reciprocal lattice point (Fig.1b). When the displacements are periodic, the two shifted Bragg peaks are not observed directly. Interference between the two reciprocal lattices replaces them with a ladder of satellite peaks pinned to the reflection of the reference structure, as in a nanotwin superlattices [35]. The satellite positions are fixed by the stripe period, and their intensity depends on the relative shift, $\mathbf{\Delta_G}$, between the two reciprocal lattices (Supplement). In thin films, additional interference fringes, Laue fringes, appear due to the finite thickness. When $\mathbf{G}$ is perpendicular to displacement $\mathbf{l}$ ($024$ peak in Fig.1c), $\mathbf{\Delta_G}$ vanishes and only vertical thickness fringes appear. When $\mathbf{G}$ is nearly parallel to $\mathbf{l}$ ($0\bar{2}4$ peak in Fig.1c), $\mathbf{\Delta_G}$ is large and intense diagonal satellites decorate the reflection in addition to vertical thickness fringes. The intensity of the satellites therefore encodes the Bragg-peak separation between the two reciprocal lattices (Supplement). Testing the applicability of equation (1) requires recording satellite shapes across many Bragg reflections spanning a range of projections $\mathbf{G} \cdot \mathbf{l}$.

### *Reciprocal space mapping and satellite peak analysis*

We investigate a 17-nm thick $Ca_2RuO_4$ film which was grown on a (001)-oriented $LaAlO_3$ substrate (pseudocubic notation) by molecular-beam epitaxy [9,24,36]. To experimentally determine the shape of the satellite peaks around multiple reciprocal lattice points, we collected a large reciprocal space volume. A parallel monochromatic X-ray beam with an energy of 15 keV illuminated the $Ca_2RuO_4$ thin film (cooled to 50 K using a cryojet) at near-grazing incidence, while an area detector captured the diffracted photons (Fig.1d). By rotating the sample azimuthally by 360°, we collected a series of 3600 diffraction images, each corresponding to a slice through reciprocal space on the Ewald sphere. We calculated the reciprocal space coordinates for each detector pixel and azimuthal angle and reconstructed the full 3D reciprocal space of the thin film via interpolation (Supplement). The resulting reciprocal space continuously spans a volume of approximately $(5\ \text{Å}^{-1})^3$ containing hundreds of reciprocal lattice points of the $Ca_2RuO_4$ structure. The high resolution of the reciprocal space scan (of order 0.01 $\text{Å}^{-1}$) allows us to resolve satellite peaks around each reciprocal lattice point (Fig.1e, see Fig. S1 for more slices through reciprocal space).

Below the metal-to-insulator transition the structure organizes into striped domains, with alternating structures resembling the L-*Pbca* and S-*Pbca* bulk structure [9]. In a cut perpendicular to the stripes, two crystallographically equivalent interfaces form [9,23]. The x-ray interference between these interfaces generates two diagonal satellite streaks visible in the vicinity of the $206$ reciprocal lattice point (Fig.2a; the inset schematic shows the interfaces perpendicular to the stripes). The satellites are narrow along *h* in reciprocal space, implying that the stripes extend over many unit cells in the [100] direction. The full reciprocal-space dataset further shows that the satellite shape in the *k*–*l* plane is independent of *h*, revealing that the displacements across the interfaces lie entirely within the $(100)$ plane. The IPS framework (Eq.1, see supplement) predicts that the satellite shape depends on the projection of the reciprocal lattice vector, $\mathbf{G}$, onto the relative displacements inside the domains, $\mathbf{l}$. Figure 2a shows six representative reflections. Indeed, as $\mathbf{G}$ rotates in the *k*–*l* plane, the two satellite streaks alternate between equal intensity (e.g., $131$, $206$) and selective extinction of one variant (e.g., $2\bar{2}6$, 226), in agreement with Eq. (1). The schematics in Figure 2a show the interface variants visible in each peak. Figure 2b shows the reciprocal space across the *k*–*l* plane in presence of two equivalent interface variants and qualitatively reproduces the appearance of satellite peaks across the whole measured reciprocal space.

To quantify the crystallographic displacements in the neighboring domains across the interfaces, we define the Symmetry Quotient as $\mathrm{SQ}_{hkl} = \min(I_1, I_2)/\max(\mathrm{I}_1, \mathrm{I}_2)$, where $I_1$ and $I_2$ are the integrated intensities of the two satellite streaks around a given Bragg reflection $hkl$. This quantity equals 1 when the satellite pattern is symmetric and both streaks are equally bright and 0 when only one streak is present. We experimentally determine the Symmetry Quotient on 24 symmetry-inequivalent Bragg reflections and plot it as a function of $\theta$, the angle between the projection of the reciprocal lattice vector $\boldsymbol{G}$ onto the *k–l* plane and sample normal [001] (Fig. 2c, black dots). The data reveal maxima near $\theta = 0$ and $\theta = \pm 90°$, where both streaks contribute equally, and minima in between. We develop an analytic formalism that quantitatively describes the intensity of satellites: when the shift of the reciprocal lattice point is small, the satellite intensity for a single interface orientation is proportional to $I_{1,2} \propto \left|\boldsymbol{G} \cdot \boldsymbol{l}_{\mathbf{1,2}}\right|^2 \propto \cos^2(\theta_{1,2})$ (Supplement). The curve calculated assuming the interfaces are $(012)$ and $(0\bar{1}2)$ (angled at $\arctan(2b/c) \approx 42°$ with respect to $(001)$) [9,24] and displacements are parallel to the interfaces closely follows the experimental data across the full angular range. This parameter-free agreement establishes the invariant plane strain framework as the governing description of the nanotexture with coherent interfaces and the inter-domain displacements along $[01\bar{2}]$ and $[0\bar{1}\bar{2}]$. Cryogenic scanning transmission electron microscopy shows no defects across the phase boundary, consistent with coherent interfaces (Fig.S3).

### *Symmetry inside the coexisting phases*

Having described the satellite patterns quantitatively, we can now use the reciprocal space data to investigate the crystal structure within the stripes of alternating distinct phases. We first discuss the space group of the parent structure by investigating the Bragg peak intensities. The films are commensurately strained (the crystal truncation rod from the substrate (top) aligns with the film peak, see $2\bar{2}6$, $226$ peaks in Fig. 2b, Fig. S4), and the pseudocubic metric imposed by the substrate may induce a change to a tetragonal space group [37], as observed in orthorhombic $CaRuO_3$ grown on cubic substrates [38]. Figure 3(a) shows crystal truncation rods extracted from the 3D reciprocal-space maps at 50 K, qualitatively similar to measurements above room temperature (Fig. S4). These data show sharp substrate peaks and broader film peaks decorated with Laue fringes. In addition, the crystal truncation rods reveal intensity at positions that are structurally forbidden in the *Pbca* space group (Table S2), e.g., the $205$ reflection. At each forbidden reciprocal lattice point *hkl*, intensity is observed only if the corresponding *khl* peak (*k* and *h* swapped) is allowed by the *Pbca* extinction rules. For example, the *012* peak is forbidden, but the *102* peak is allowed, and thus intensity appears at the *012* position because of twinning. At positions where both *hkl* and *khl* are forbidden by different glide plane symmetries, such as *103/013* and *105/015*, no intensity is observed. This rules out a symmetry change and confirms that the intensity at reflections forbidden in the orthorhombic space group originates from different twin domains rotated by 90° with respect to each other around the film normal.

Two orthogonal stripe patterns coexist in macroscopically distinct regions of the film, generating satellites with a cross shape when viewed in a fixed-*l* slice [9,24] (Fig. 3b). This shape has two possible origins: the nanodomains may extend along both in-plane axes of the unit cell, or along a single axis and merely appear doubled because of the orthorhombic twins discussed above. The Bragg peaks alone cannot distinguish the two, since 90° twins overlap. The satellites resolve this ambiguity, because the reciprocal-lattice shift $\mathbf{\Delta_G}$ between the two coexisting structures and the satellite peaks orient along the interface normal $\boldsymbol{n}$ (Eq.1). At the 206 and 026 positions, where the reflection and its orthorhombic twin are both allowed, the satellites form a cross of two perpendicular satellite patterns (Fig. 3b, top row). At the 104 position, where 104 is allowed but 014 is forbidden, only one satellite orientation appears (none forms along the direction of the absent 014 peak, Fig. 3b, bottom row). The stripes therefore extend along a single crystallographic direction, [100]. Figure 3c shows the resulting reciprocal-lattice schematic for fixed-*l* slices at odd and even **l**: the allowed peaks and satellite streaks from the primary domain in red, the 90°-rotated orthorhombic twin in blue.

Finally, the satellites enable us to test the space group of the coexisting structures in the stripes. Each satellite pattern carries contributions from the structure factors of both coexisting structures (Supplement). A satellite therefore has non-zero intensity only where the displacement projection $\mathbf{G} \cdot \mathbf{l}$ is non-zero and at least one of the two domains contributes an allowed Bragg reflection. Among reflections with $\mathbf{G} \cdot \mathbf{l} \neq 0$, the absence of satellites thus implies that the reflection is forbidden in both domain structures. We find that satellites are absent at every such position forbidden by the *Pbca* reflection conditions, e.g., 014 (Table S2). Both structures forming the stripe pattern therefore obey the same reflection conditions as the bulk *Pbca* phase. This observation adds to the persistence of *Pbca* in bulk under several-percent uniaxial strain [39] and moderate hydrostatic pressure [19].

## *Discussion and Conclusions*

Using large-volume X-ray reciprocal-space mapping, we recorded diffraction intensity in the vicinity of hundreds of reciprocal-lattice points of an epitaxially strained $Ca_2RuO_4$ thin film. Below the metal-insulator transition, the striped domain structure decorates the Bragg reflections with satellite peaks. We derived from geometric martensitic theory a closed-form scattering model in which the satellite intensity scales as $|\mathbf{G} \cdot \mathbf{l}|^2$. Within this model, the measured satellite intensities across 24 symmetry-inequivalent reflections collapse onto a single parameter-free curve. The collapse identifies the striped state as a coherent martensitic laminate of few-nm-wide domains separated by $(012)$ and $(0\bar{1}2)$ interfaces, with inter-domain displacements along the $[01\bar{2}]$ and $[0\bar{1}\bar{2}]$ direction. Satellite-extinction analysis shows that both coexisting phases, while differing in lattice constants, share the bulk orthorhombic *Pbca* space group despite biaxial epitaxial strain from the substrate and transformation strain at the interfaces. In this biaxially strained Mott insulator film, the striped domain geometry is therefore predominantly set by classical structural strain compatibility, with no strong contributions from electronic or magnetic order.

In bulk $Ca_2RuO_4$, all three lattice parameters change across the phase transformation, complicating structural compatibility between the L and S phases along a coherent interface. The substrate epitaxy restricts expansion along [100], so only **b** and **c** change across the transition. This restriction guarantees the existence of a plane with ideal structural compatibility: $(012)$ and $(0\bar{1}2)$) [31]. Lattice compatibility was shown to control the metal-insulator transition hysteresis in the correlated oxide $V_{1-x}W_xO_2$ [40]; here, epitaxial clamping instead enforces a compatible plane that templates the periodic laminate. Coherent, structurally compatible interfaces are required for reproducible structural transformations and low fatigue upon cycling [41]. Epitaxial strain may therefore serve as a tool to engineer this coherency by enforcing existence of coherent interfaces.

The structural stability of the $Ca_2RuO_4$ thin films through thousands of optically induced transitions in stroboscopic pump-probe measurements is consistent with this picture [36].

Large-volume reciprocal-space mapping combined with invariant-plane-strain analysis thus resolves the structure of coexisting domains from a single multi-reflection diffraction dataset. Because the framework requires only geometric inputs, the same parameter-free analysis can resolve the crystallography of coexisting domains in heteroepitaxial nickelates [42], vanadates [43], and ferroelastic twins in epitaxial ferroelectric films [44]. The diffraction measurement is non-invasive, and the analysis extends naturally to time-resolved studies of domain-wall motion in ferroics under electric fields [44] and of the picosecond reorganization of the striped Mott state in $Ca_2RuO_4$ under current [25] and light [36].

***Acknowledgements***

The authors acknowledge the significant contribution of the late Professor Lena F. Kourkoutis to this study. This work was primarily supported by the Department of Energy – Office of Basic Energy Sciences under award DE-SC0019414 (B.Z.G, Y.A.B., Z.S., J.H, K.M.S., D.G.S., A.S., thin film growth, reciprocal space mapping, X-ray diffraction model). This research was funded in part by the Gordon and Betty Moore Foundation's EPiQS Initiative (Grant Nos. GBMF3850 and GBMF9073 to Cornell University). Research conducted at the Center for High-Energy X-ray Sciences (CHEXS) is supported by the National Science Foundation (BIO, ENG and MPS Directorates) under award DMR-2342336.

## Figures

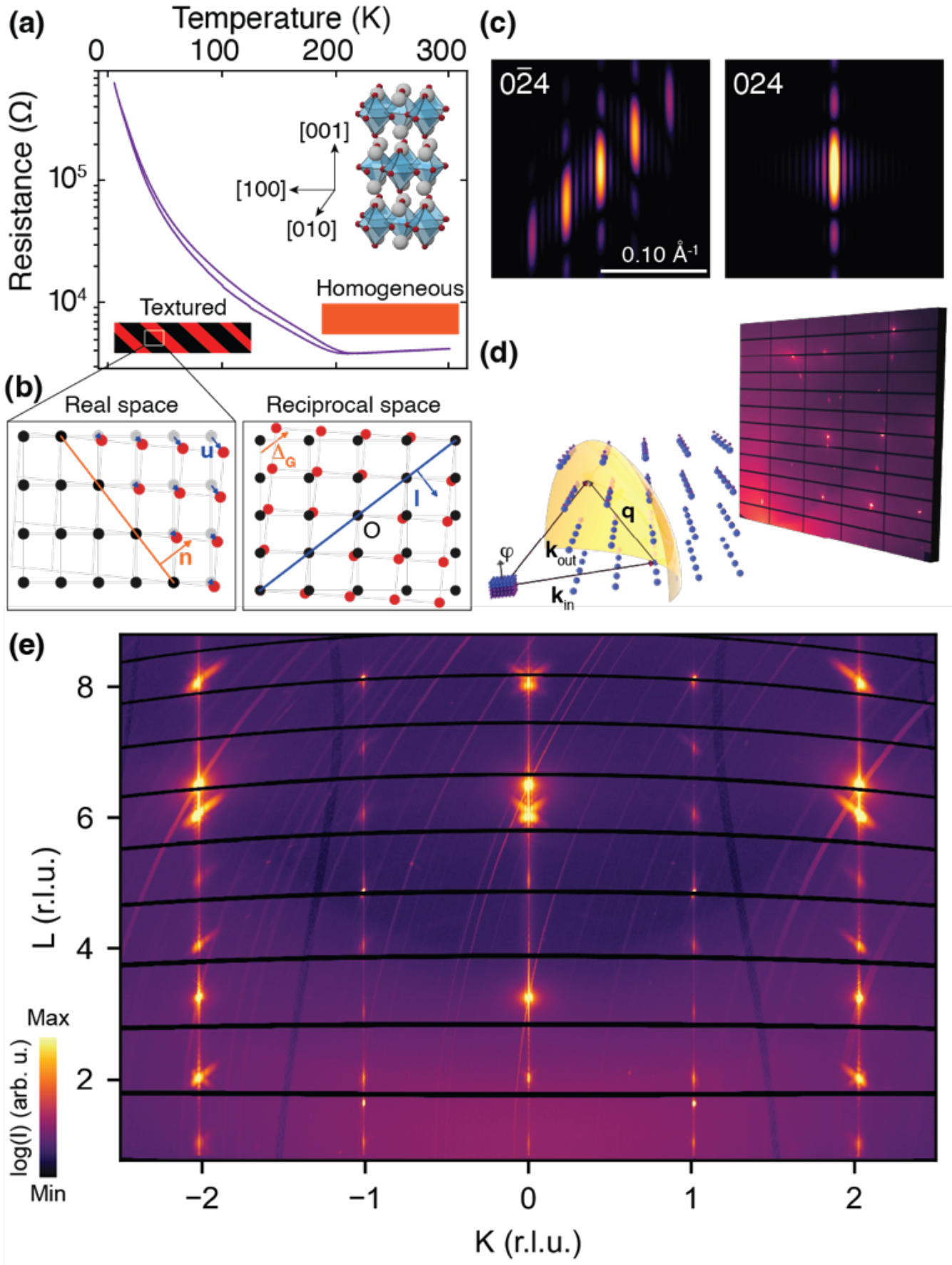

**Figure 1: Striped nanotexture and reciprocal space mapping of $Ca_2RuO_4$.** (a) Resistivity of the strained $Ca_2RuO_4$ thin film showing a metal-insulator transition at 200 K. The film is homogeneous at high temperatures and forms a periodic nanotexture of distinct structural domains resembling the bulk L-*Pbca* and S-*Pbca* phases (red and black stripes). (Inset) $Ca_2RuO_4$ unit cell (Ca: gray, O: red, Ru: blue). (b) Real space: atomic positions in the two structural variants, with interface normal $\boldsymbol{n}$ and the displacements $\boldsymbol{u}$ all parallel to $\mathbf{l}$. Reciprocal space: The reciprocal lattice points of the distorted lattice are shifted by $\boldsymbol{\Delta}_\mathbf{G}$ along $\mathbf{n}$. The lattice remains undistorted along the line through the origin normal to $\mathbf{l}$. (c) Simulated satellite patterns for a periodic stripe domain with the invariant plane (012) and displacement confined to that plane ($|\boldsymbol{\Delta}_{0\bar{2}4}| > 0$; $\boldsymbol{\Delta}_{024} = 0$). (d) Experimental schematic: $\mathbf{k}_{\mathrm{in}}$ and $\mathbf{k}_{\mathrm{out}}$ are wavevectors of the incoming and diffracted x-ray beams, and their difference is the reciprocal space vector $\mathbf{q}$. Rotation of the sample around the film normal maps out reciprocal space on an area detector. (e) Diffracted intensity on a slice through the reciprocal lattice at $h = 2$ on a logarithmic scale measured at 50 K.

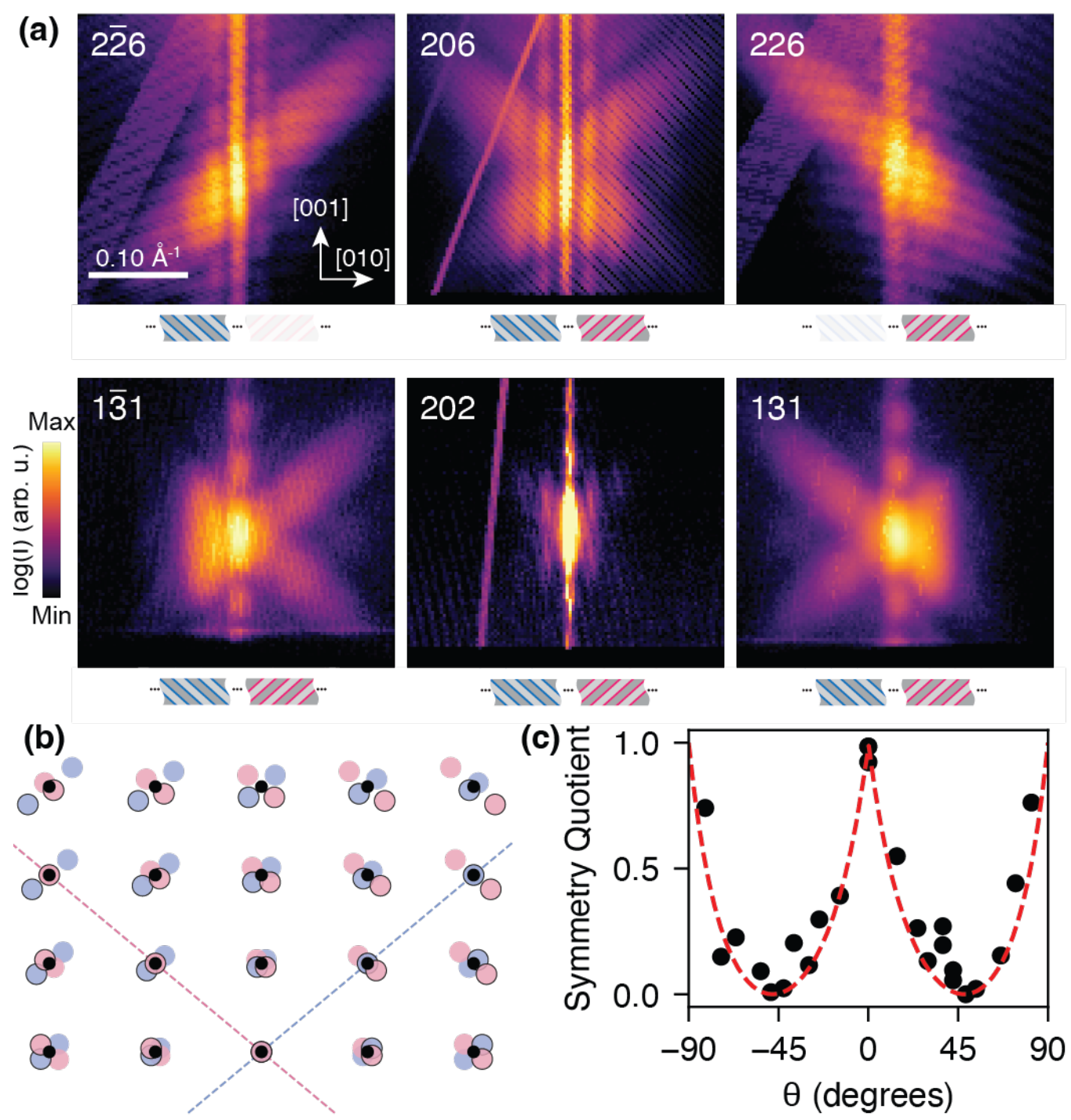

**Figure 2: Parameter-free collapse of satellite intensities onto the invariant-plane-strain prediction.** (a) Magnified reciprocal space maps around selected reciprocal lattice points. The schematics under each panel show which interface variant contributes to the satellite pattern at that reflection (interface along (012) (blue), or along (0$\bar{1}$2) (red), or both). (b) Reciprocal-space schematic showing the contributions of both interface variants. Symbols denote the two coexisting structures (S-*Pbca*, no edge color; L-*Pbca*, black edge). Dashed lines are normal to the displacement direction for each interface variant. (c) Symmetry quotient measured at 24 symmetry-inequivalent reflections with sufficient satellite intensity, plotted versus the angle between reciprocal lattice point *hkl* and film normal [001] (black dots). The red dashed curve is the parameter-free prediction of the invariant-plane-strain model (Supplement).

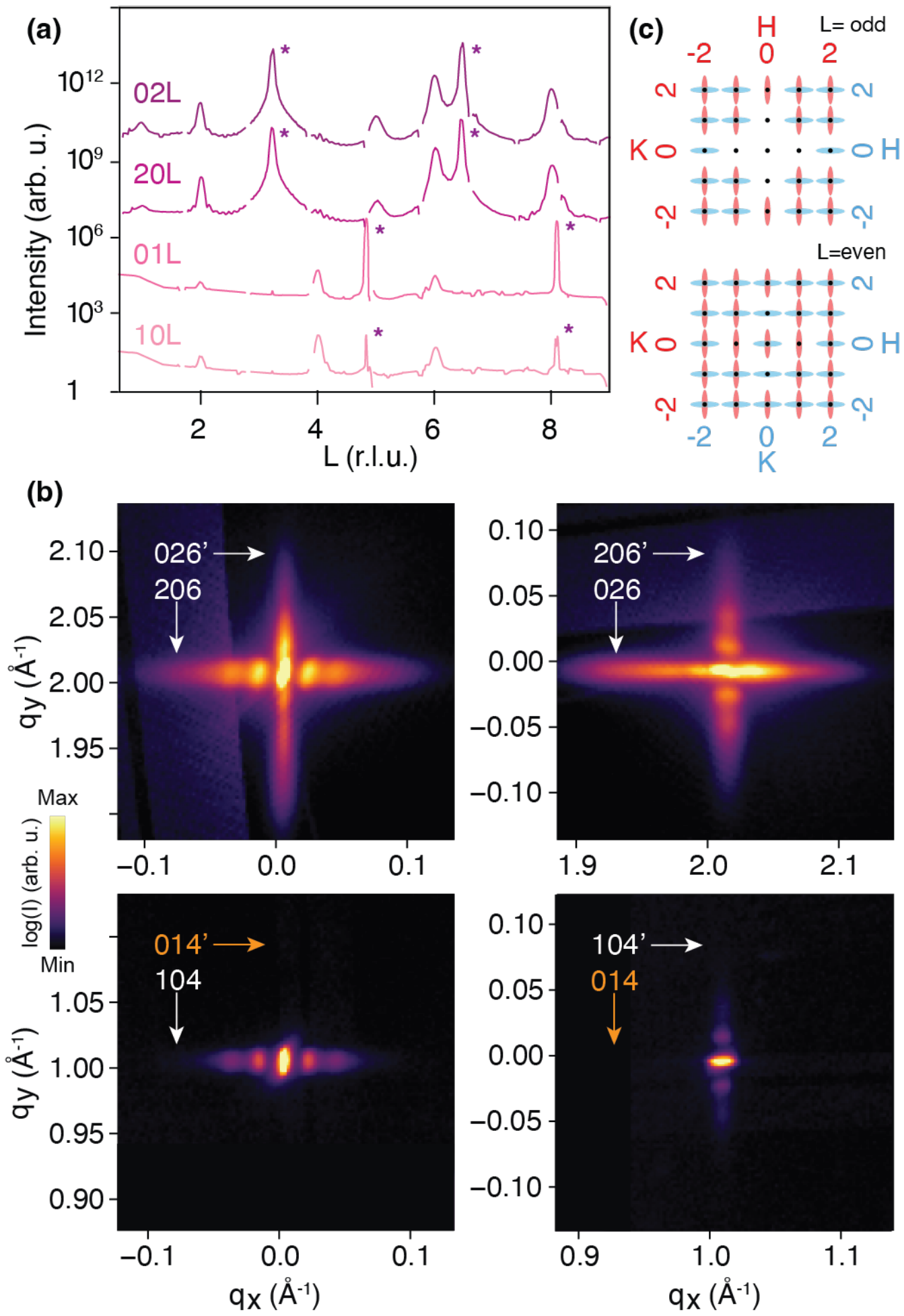

**Figure 3: Crystallography of coexisting domains.** (a) Crystal truncation rods extracted from the 3D reciprocal-space map of $Ca_2RuO_4$ at 50 K, taken along **l** at four in-plane reciprocal-lattice positions *hk* = 02, 20, 01, and 10; asterisks mark peaks from the $LaAlO_3$ substrate. At 300 K, the peak intensities are similar, and satellites are absent (Fig. S4). (b) Satellite patterns around the indicated reflections, shown in slices through reciprocal space at fixed *l*. Unprimed indices label the primary orthorhombic twin; primed indices label the 90°-rotated twin. At 206 and 026 (top row), both reflections are *Pbca*-allowed, and the satellites from the two variants form a cross. At 104 and 014 (bottom row), only one reflection is *Pbca*-allowed in each pair (014 is forbidden) and satellites appear in a single direction. (c) Schematic of allowed reflections and the associated satellite patterns. The primary domain (blue) produces satellites along *k*; the 90°-rotated twin (red) produces satellites along *h*. The patterns differ for *l*=odd (top) and *l*=even (bottom) according to the *Pbca* extinction rules.

## References

[1] L. Landau and E. Lifshits, On the theory of the dispersion of magnetic permeability in ferromagnetic bodies, Physik. Zeits. Sowjetunion **8**, 153 (1935).

[2] C. Kittel, Theory of the Structure of Ferromagnetic Domains in Films and Small Particles, Phys. Rev. **70**, 965 (1946).

[3] T. Mitsui and J. Furuichi, Domain Structure of Rochelle Salt and KH2PO4, Phys. Rev. **90**, 193 (1953).

[4] A. L. Roitburd, Equilibrium structure of epitaxial layers, Phys. Stat. Sol. (a) **37**, 329 (1976).

[5] G. Catalan, J. F. Scott, A. Schilling, and J. M. Gregg, Wall thickness dependence of the scaling law for ferroic stripe domains, J. Phys.: Condens. Matter **19**, 022201 (2007).

[6] M. Uehara, S. Mori, C. H. Chen, and S.-W. Cheong, Percolative phase separation underlies colossal magnetoresistance in mixed-valent manganites, Nature **399**, 560 (1999).

[7] A. S. McLeod et al., Nanotextured phase coexistence in the correlated insulator V2O3, Nature Physics **13**, 80 (2016).

[8] A. Singer et al., Nonequilibrium Phase Precursors during a Photoexcited Insulator-to-Metal Transition in V2O3, Phys Rev Lett **120**, 207601 (2018).

[9] Z. Shao et al., Real-space imaging of periodic nanotextures in thin films via phasing of diffraction data, Proc Natl Acad Sci U S A **120**, e2303312120 (2023).

[10] V. Podzorov, B. G. Kim, V. Kiryukhin, M. E. Gershenson, and S.-W. Cheong, Martensitic accommodation strain and the metal-insulator transition in manganites, Phys. Rev. B **64**, 140406 (2001).

[11] K. H. Ahn, T. Lookman, and A. R. Bishop, Strain-induced metal–insulator phase coexistence in perovskite manganites, Nature **428**, 401 (2004).

[12] E. Dagotto, Complexity in strongly correlated electronic systems, Science **309**, 257 (2005).

[13] M. Braden, G. André, S. Nakatsuji, and Y. Maeno, Crystal and magnetic structure ofCa2RuO4:Magnetoelastic coupling and the metal-insulator transition, Physical Review B **58**, 847 (1998).

[14] C. S. Alexander, G. Cao, V. Dobrosavljevic, S. McCall, J. E. Crow, E. Lochner, and R. P. Guertin, Destruction of the Mott insulating ground state of Ca2RuO4 by a structural transition, Physical Review B **60**, R8422 (1999).

[15] O. Friedt, M. Braden, G. André, P. Adelmann, S. Nakatsuji, and Y. Maeno, Structural and magnetic aspects of the metal-insulator transition in Ca2−xSrxRuO4, Physical Review B **63**, 174432 (2001).

[16] E. Gorelov, M. Karolak, T. O. Wehling, F. Lechermann, A. I. Lichtenstein, and E. Pavarini, Nature of the Mott transition in Ca2RuO4, Phys Rev Lett **104**, 226401 (2010).

[17] Q. Han and A. Millis, Lattice Energetics and Correlation-Driven Metal-Insulator Transitions: The Case of $Ca_2RuO_4$, Physical Review Letters **121**, 067601 (2018).

[18] F. Nakamura, M. Sakaki, Y. Yamanaka, S. Tamaru, T. Suzuki, and Y. Maeno, Electric-field-induced metal maintained by current of the Mott insulator Ca2RuO4, Sci Rep **3**, 2536 (2013).

[19] P. Steffens et al., High-pressure diffraction studies on Ca2RuO4, Physical Review B **72**, 094104 (2005).

[20] L. Miao, P. Silwal, X. Zhou, I. Stern, J. Peng, W. Zhang, L. Spinu, Z. Mao, and D. Ho Kim, Itinerant ferromagnetism and geometrically suppressed metal-insulator transition in epitaxial thin films of Ca2RuO4, Applied Physics Letters **100**, (2012).

[21] C. Dietl et al., Tailoring the electronic properties of Ca2RuO4 via epitaxial strain, Applied Physics Letters **112**, (2018).
[22] A. Tsurumaki-Fukuchi, K. Tsubaki, T. Katase, T. Kamiya, M. Arita, and Y. Takahashi, Stable and Tunable Current-Induced Phase Transition in Epitaxial Thin Films of $Ca_2 RuO_4$, ACS Appl. Mater. Interfaces **12**, 28368 (2020).
[23] N. Schnitzer, G. Powers, B. H. Goodge, E. Bianco, I. E. Baggari, and L. F. Kourkoutis, Atomic-Resolution Imaging of Phase Transitions in Strongly Correlated Oxides with Continuously Variable Temperature Cryo-STEM, Microscopy and Microanalysis **29**, 1688 (2023).
[24] O. Y. Gorobtsov et al., Spontaneous Supercrystal Formation During a Strain-Engineered Metal-Insulator Transition, Adv Mater **36**, e2403873 (2024).
[25] J. Zhang et al., Nano-Resolved Current-Induced Insulator-Metal Transition in the Mott Insulator Ca2RuO4, Physical Review X **9**, 011032 (2019).
[26] K. Jenni, F. Wirth, K. Dietrich, L. Berger, Y. Sidis, S. Kunkemöller, C. P. Grams, D. I. Khomskii, J. Hemberger, and M. Braden, Evidence for current-induced phase coexistence in Ca2RuO4 and its influence on magnetic order, Physical Review Materials **4**, (2020).
[27] J. Bang, N. Strkalj, M. F. Sarott, Y. Kholina, M. Trassin, and T. Weber, High-energy diffuse X-ray scattering at ultra-small-angle grazing incidence for local structure study of single-crystalline thin films, J Appl Crystallogr **58**, 1417 (2025).
[28] Y. Tao, E. Peng, Q. He, X. Chen, J. Li, Y. Wang, Y. Dong, Z. Sun, and Z. Luo, Simulation of three dimensional diffraction patterns as aid of structural analysis for complex epitaxial films, Sci Rep **15**, 18328 (2025).
[29] E. Zatterin et al., Assessing the Ubiquity of Bloch Domain Walls in Ferroelectric Lead Titanate Superlattices, Phys. Rev. X **14**, 041052 (2024).
[30] V. A. Stoica et al., Non-equilibrium pathways to emergent polar supertextures, Nat Mater **23**, 1394 (2024).
[31] A. Khachaturyan, *Theory of Structural Transformations in Solids*, 1st ed. (Dover Publications, 2008).
[32] M. S. Wechsler, D. S. Lieberman, and T. A. Read, On the theory of the formation of martensite, Trans. Met. Soc. AIME **197**, 1503 (1953).
[33] J. S. Bowles and J. K. Mackenzie, The crystallography of martensite transformations I, Acta Metallurgica **2**, 129 (1954).
[34] J. Sapriel, Domain-wall orientations in ferroelastics, Phys. Rev. B **12**, 5128 (1975).
[35] Y. U. Wang, Diffraction theory of nanotwin superlattices with low symmetry phase, Phys. Rev. B **74**, 104109 (2006).
[36] A. Verma et al., Picosecond volume expansion drives a later-time insulator-metal transition in a nano-textured Mott insulator, Nature Physics **20**, 807 (2024).
[37] J. M. Rondinelli and N. A. Spaldin, Structure and properties of functional oxide thin films: insights from electronic-structure calculations, Adv Mater **23**, 3363 (2011).
[38] A. Vailionis, W. Siemons, and G. Koster, Room temperature epitaxial stabilization of a tetragonal phase in ARuO3 (A=Ca and Sr) thin films, Applied Physics Letters **93**, 051909 (2008).
[39] S. Riccò et al., In situ strain tuning of the metal-insulator-transition of Ca2RuO4 in angle-resolved photoemission experiments, Nature Communications **9**, 4535 (2018).
[40] Y. G. Liang et al., Tuning the hysteresis of a metal-insulator transition via lattice compatibility, Nat Commun **11**, 3539 (2020).

[41] C. Chluba, W. Ge, R. Lima De Miranda, J. Strobel, L. Kienle, E. Quandt, and M. Wuttig, Ultralow-fatigue shape memory alloy films, Science **348**, 1004 (2015).
[42] G. Mattoni et al., Striped nanoscale phase separation at the metal–insulator transition of heteroepitaxial nickelates, Nat Commun **7**, 13141 (2016).
[43] L. Rodríguez, F. Sandiumenge, C. Frontera, J. M. Caicedo, J. Padilla, G. Catalán, and J. Santiso, Strong strain gradients and phase coexistence at the metal-insulator transition in VO2 epitaxial films, Acta Materialia **220**, 117336 (2021).
[44] G. Catalan, J. Seidel, R. Ramesh, and J. F. Scott, Domain wall nanoelectronics, Rev. Mod. Phys. **84**, 119 (2012).

**Supplementary Information**

# Crystallography of periodic nanotextures in a strained Mott insulator

*Benjamin Z. Gregory*[1,2], *Yorick A. Birkhölzer*[1], *Noah Schnitzer*[1], *Ziming Shao*[1], *Jeff Hodgson*[1], *Suchismita Sarker*[3], *Jacob P. Ruff*[3], *Berit H. Goodge*[4], *David A. Muller*[5,6], *Kyle M. Shen*[2,6], *Darrell G. Schlom*[1,6,7], *Andrej Singer*[1,*]

[1]*Department of Materials Science and Engineering, Cornell University; Ithaca, NY 14853, USA*
[2]*Department of Physics, Cornell University; Ithaca, NY 14853, USA*
[3]*Cornell High Energy Synchrotron Source, Cornell University; Ithaca, NY 14853*
[4]*Max Planck Institute for Chemical Physics of Solids; 01187 Dresden, Germany*
[5] *School of Applied and Engineering Physics, Cornell University; Ithaca, NY 14853, USA*
[6]*Kavli Institute at Cornell for Nanoscale Science, Cornell University; Ithaca, NY 14853, USA*
[7]*Leibniz-Institut für Kristallzüchtung; Max-Born-Straße 2, Berlin 12489, Germany*
[*]*asinger@cornell.edu*

| | S-Pbca (180 K) | L-Pbca (400 K) | $(d_S - d_L)/d_L$ |
|---|---|---|---|
| a (Å) | 5.3945 | 5.3606 | +0.6 % |
| b (Å) | 5.5999 | 5.3507 | +4.7% |
| c (Å) | 11.7653 | 12.2637 | -4.1% |
| | S-Pbca (357 K) | L-Pbca (357 K) | $(d_S - d_L)/d_L$ |
| a (Å) | 5.40 | 5.37 | +0.6% |
| b (Å) | 5.44 | 5.36 | +1.5% |
| c (Å) | 12.08 | 12.26 | -1.5% |
| | S-Pbca (300K) | $LaAlO_3$ | $(d_{LAO} - d_S)/d_S$ |
| a (Å) | 5.41 | 5.36 | -0.9% |
| b (Å) | 5.49 | 5.36 | -2.4% |

**Table S1:** Lattice parameters for bulk $Ca_2RuO_4$ reported in ref. [1]. The values for the S-Pbca (180 K) and L-Pbca (400K) phases are taken directly from tabulated data. The values at 357 K were estimated from published figures. For $LaAlO_3$, we used the pseudocubic lattice constant $a_{pc} = 3.79$Å. Assuming epitaxial alignment of the pseudocubic [110] direction of $LaAlO_3$ with the [100] direction of orthorhombic $Ca_2RuO_4$, the effective in plane substrate lattice parameter is $a = b = \sqrt{2}\ a_{pc} = 5.36$Å. At 300 K, the geometric mean of the in-plane strain imposed by the $LaAlO_3$ substrate on the S-Pbca phase, which is stable in bulk at 300K, is $-\sqrt{0.9 \times 2.4}\% = -1.5\%$.

| Reflection | Condition |
|---|---|
| 0kl | k = 2n |
| h0l | l = 2n |
| hk0 | h = 2n |
| h00 | h = 2n |
| 0k0 | k = 2n |
| 00l | l = 2n |

**Table S2.** Reflection conditions for *Pbca*: Bragg peak is present if condition is satisfied.

## Reciprocal-space mapping: data processing and analysis pipeline

The experimental geometry is shown in Fig. 1d. An unfocused 15 keV X-ray beam is incident on the film at a shallow angle of about 4°, chosen such that the beam footprint does not exceed the 5-10 mm film size. Diffraction data are recorded on a Pilatus 6M area detector while the film is rotated continuously around its surface normal. The sample to detector distance is about 0.5 m and the detector is positioned such that the direct beam is close to the detector corner, allowing maximum extent in reciprocal space. A full 360° azimuthal scan, $\varphi$-scan, acquires 3600 images in 10 minutes. To close the gaps between detector panels, the measurement is repeated at two incident angles $\theta$; this pair is repeated at two $\chi$ angles, which tilt the film normal a few degrees out of the scattering plane. A separate $\theta$-scan is collected to fill in the specular reflections, which are missing in the azimuthal scan at a fixed $\theta$. The complete dataset comprises four $\varphi$ scans and 30 minutes of acquisition.

Each detector pixel at a given $(\theta, \chi, \varphi)$ corresponds to a reciprocal-space point $\mathbf{Q}_{\mathrm{pix}}$ on the Ewald sphere. Following Busing and Levy [2], the corresponding point in the reciprocal lattice of the crystal is given by

$$\mathbf{h} = (\mathbf{UB})^{-1}\mathbf{R}_{\mathrm{gonio}}^{-1}\mathbf{Q}_{\mathrm{pix}}, \qquad (1)$$

where $\mathbf{R}_{\mathrm{gonio}}$ rotates $\mathbf{Q}_{\mathrm{pix}}$ into the lab frame $\mathbf{Q}_{\mathrm{lab}}$ (defined here by $\theta = \chi = \varphi = 0$) and the $\mathbf{U}$ is a rotation matrix, and $\mathbf{B}$ maps reciprocal space onto the crystal reciprocal lattice (in reciprocal-lattice units).

Detector calibration is performed once per beamtime against a powder calibrant ($CeO_2$) using the Python package pyFAI [3], which returns a PONI file with the six parameters describing the detector position and orientation in 3D space. Malfunctioning pixels, panel gaps, and strong small-angle scattering near the direct beam are marked in a mask file, and are excluded from the analysis. With the PONI file, pyFAI assigns to every pixel a unique transfer wavevector $\mathbf{Q}_{\mathrm{pix}}$ in the lab Cartesian basis. The goniometer rotation matrix $\mathbf{R}_{\mathrm{gonio}}$ is constructed from the Euler angles using SciPy package [4]. We implement Eq. (1) using two approaches (Fig. S5). The two differ in when the coordinate transform from detector pixels to reciprocal space is performed and in how many interpolation steps are required to obtain a particular view of the data. Critically, the first approach is agnostic to the crystal structure under investigation and allows on-the-fly analysis simultaneous with data collection. The second approach is only suitable for post-experiment analysis, as it requires adjustments for each specific crystal structure.

The **first approach**, used here to obtain the large-volume reciprocal-space dataset, proceeds in three stages: a one-time conversion of the raw diffraction images into a 3D reciprocal-space volume, a fast extraction of the crystal orientation from that volume, and rapid (within seconds) extraction of arbitrary 2D slices or 1D line cuts.

*Conversion to a 3D reciprocal-space volume.* Raw diffraction images are mapped into the lab frame via $\mathbf{Q}_{\mathrm{lab}} = \mathbf{R}_{\mathrm{gonio}}^{-1}\mathbf{Q}_{\mathrm{pix}}$ and interpolated onto a 3D Cartesian grid ($\mathbf{U} = \mathbf{1}$). We do not apply the $\mathbf{B}$ matrix at this stage; the volume is expressed in lab-frame Å$^{-1}$, not in reciprocal-lattice units. This step is computationally heavy and takes about twenty minutes per dataset.

*Crystal orientation from the 3D volume.* To determine the orientation, we locate ~50 of the strongest peaks in $\mathbf{Q}_{\text{lab}}$ — typically substrate reflections — and compute all possible directions between all found reciprocal lattice points. Given the substrate unit cell as input, we use the scalar triple product of three measured vectors as a rotation-invariant matching criterion: $\mathbf{Q}_1 \cdot (\mathbf{Q}_2 \times \mathbf{Q}_3)$ equals the parallelepiped volume formed by the three vectors and is conserved under any orientation of the crystal. By comparing the triple products of measured peak triplets to those of all triplets of known substrate reciprocal-lattice vectors, we identify the assignment of measured directions to lattice vectors that best matches the unit-cell geometry. This assignment determines the orientation matrix **U**. This step takes seconds.

*Display and analysis of the reciprocal space data.* With **U** determined, the 3D volume is interpolated onto an arbitrary 2D plane, an arbitrary 1D line, or a 3D subset along directions of interest in lab-frame $Å^{-1}$. Optionally, the **B** matrix (built from the known unit cell) can be applied during this step to express the output in reciprocal-lattice units, and additional rotations can transform between crystal frames — for example, from pseudocubic $LaAlO_3$ to orthorhombic *Pbca* by a 45° rotation around the *c*-axis. This second interpolation is fast, taking seconds, and can be repeated arbitrarily many times across the cached 3D volume.

The full pipeline compresses ~100 GB of detector images into a ~2 GB 3D reciprocal-space dataset spanning $(5\ Å^{-1})^3$ on a 1000×1000×1000 grid with a voxel size of 0.005 $Å^{-1}$ (the resolution is of order 0.01 $Å^{-1}$ or better dependent on the position of the detector pixel). The first step runs about as fast as the data acquisition itself and can therefore execute concurrently with measurement to deliver the 3D reciprocal-space map shortly after a scan completes. Once cached, any number of slices and cuts can be generated within seconds. The approach is well-suited to high-throughput experiments and to on-the-fly data analysis at the beamline, for example by automatically transforming the data to 3D reciprocal space once the data are recorded, as done in this experiment.

The **second approach** is used when the highest reciprocal-space resolution is required around specific peaks. The data are mapped directly from the raw images onto a 3D grid aligned with the crystal axes of the film using the **UB** matrix — a single interpolation step. The **UB** matrix can be obtained in two ways: from the orientation **U** recovered in the first approach (combined with the known substrate **B**) plus the known epitaxial relationship between film and substrate, or independently by indexing the diffraction data with DIALS [5], a software suite widely used in protein crystallography that allows simultaneous indexing of two crystal lattices. Either route delivers higher reciprocal-space resolution than the first approach because only one interpolation is performed and a higher number of points can be used (the detector spans 2500 pixels in each direction, making best resolution of 5000x2500 pixels neglecting Ewald sphere curvature); the cost is that the indexing-and-transform sequence must be repeated for each region of interest, so the procedure is less suited to surveying a large volume but well-suited to high-resolution measurements around individual peaks. We use this method to extract high-resolution 3D regions of interest around individual peaks for the satellite analyses presented in the main text.

## Kinematical Scattering from a Coherent Invariant-Plane-Strain Laminate

This description follows the geometric theory of martensitic phase transformations [6] and Wang's work [7], who considered periodic twin laminates. Our contribution comes at the end, where we derive an analytic expression and show how the intensity of each satellite peak depends on the orientation of the reciprocal lattice vector relative to the direction along which the atoms are displaced.

### *Single interface*

The kinematic scattering amplitude is [8]

$$A(\mathbf{q}) = \sum_j f_j \; e^{\,i\mathbf{q}\cdot\left(\mathbf{r}_j^0+\mathbf{u}_j\right)},$$

where $\mathbf{q}$ is the momentum transfer, $\mathbf{r}_j^0$ the undisplaced atomic positions, $\mathbf{u}_j$ the displacement of atom $j$, and $f_j$ the atomic form factors; the sum runs over all atoms. An invariant plane strain (IPS) is the deformation produced by a single martensitic variant: every atom is displaced along one fixed direction $\mathbf{l}$, by an amount proportional to its distance from a fixed reference plane,

$$\mathbf{u} = \varepsilon(\mathbf{n}\cdot\mathbf{r}^0)\mathbf{l} = \varepsilon\xi\mathbf{l}, \qquad \xi \equiv \mathbf{n}\cdot\mathbf{r}^0,$$

with $\xi$ measured along the plane normal $\mathbf{n}$ (Fig. 1b, real space: plane in orange). The displacement vanishes on the reference plane and increases linearly with distance from it. Within a single domain the displaced position is $\mathbf{r}^0 + \mathbf{u} = \mathbf{F}\mathbf{r}^0$, with

$$\mathbf{F} = \mathbf{1} + \varepsilon\mathbf{l}\otimes\mathbf{n}.$$

This deformation differs from the identity by the single dyad $\mathbf{l}\otimes\mathbf{n}$. It is rank-one: it acts along the single direction $\mathbf{l}$ and varies only with position along the normal $\mathbf{n}$, leaving the reference plane undistorted ($\otimes$ denotes the dyadic product, $\mathbf{1}$ the identity).

The displacement enters the scattering only through the phase, where the deformation transfers from the atomic positions to the momentum:

$$\mathbf{q}\cdot(\mathbf{r}^0+\mathbf{u}) = \mathbf{q}\cdot(\mathbf{F}\mathbf{r}^0) = (\mathbf{F}^T\mathbf{q})\cdot\mathbf{r}^0.$$

A crystal deformed by $\mathbf{F}$ and probed at $\mathbf{q}$ therefore scatters identically to an undeformed crystal probed at $\mathbf{F}^T\mathbf{q}$, and a Bragg reflection occurs wherever $\mathbf{F}^T\mathbf{q} = \mathbf{G}$ for a parent reciprocal-lattice vector $\mathbf{G}$. Solving for $\mathbf{q}$ shows that the reflection at $\mathbf{G}$ shifts to $\mathbf{G} + \mathbf{\Delta_G}$, with [6]

$$\mathbf{\Delta_G} = -\frac{\varepsilon(\mathbf{G}\cdot\mathbf{l})}{1+\varepsilon(\mathbf{n}\cdot\mathbf{l})}\mathbf{n}.$$

The shift points along $\mathbf{n}$ and grows with $\mathbf{G}\cdot\mathbf{l}$, the projection of the reciprocal-lattice vector onto the displacement direction.

It is useful to also define a shift that depends on the probing momentum directly,

$$\mathbf{\Delta}(\mathbf{q}) \equiv -\varepsilon(\mathbf{q}\cdot\mathbf{l})\mathbf{n},$$

which lets the displacement phase be written as a simple shift of the momentum,

$$e^{i\mathbf{q}\cdot(\mathbf{r}^0+\mathbf{u})} = e^{i\mathbf{q}\cdot\mathbf{r}^0} e^{i\mathbf{q}\cdot\mathbf{u}} = e^{i(\mathbf{q}-\boldsymbol{\Delta})\cdot\mathbf{r}^0}.$$

In this form, the deformed lattice scatters as if probed at $\mathbf{q} - \boldsymbol{\Delta}(\mathbf{q})$, so the Bragg condition $\mathbf{q} - \boldsymbol{\Delta}(\mathbf{q}) = \mathbf{G}$ locates the reflection at

$$\mathbf{q}_{\text{peak}} = \mathbf{G} + \boldsymbol{\Delta}_{\mathbf{G}}, \qquad \boldsymbol{\Delta}_{\mathbf{G}} = \boldsymbol{\Delta}(\mathbf{q}_{\text{peak}}) = -\varepsilon(\mathbf{q}_{\text{peak}} \cdot \mathbf{l})\mathbf{n}.$$

The shift depends on the peak position it defines, requiring a self-consistent solution. Taking the scalar product of $\mathbf{q}_{\text{peak}} = \mathbf{G} + \boldsymbol{\Delta}_{\mathbf{G}}$ with $\mathbf{l}$, and using $\boldsymbol{\Delta}_{\mathbf{G}} \cdot \mathbf{l} = -\varepsilon(\mathbf{q}_{\text{peak}} \cdot \mathbf{l})(\mathbf{n} \cdot \mathbf{l})$,

$$\mathbf{q}_{\text{peak}} \cdot \mathbf{l} = \mathbf{G} \cdot \mathbf{l} - \varepsilon(\mathbf{q}_{\text{peak}} \cdot \mathbf{l})(\mathbf{n} \cdot \mathbf{l}).$$

Collecting the $\mathbf{q}_{\text{peak}} \cdot \mathbf{l}$ terms gives a scalar equation,

$$(\mathbf{q}_{\text{peak}} \cdot \mathbf{l})[1 + \varepsilon(\mathbf{n} \cdot \mathbf{l})] = \mathbf{G} \cdot \mathbf{l},$$

and substituting the result back into $\boldsymbol{\Delta}_{\mathbf{G}} = -\varepsilon(\mathbf{q}_{\text{peak}} \cdot \mathbf{l})\mathbf{n}$ yields as before $\boldsymbol{\Delta}_{\mathbf{G}} = -\frac{\varepsilon(\mathbf{G}\cdot\mathbf{l})}{1+\varepsilon(\mathbf{n}\cdot\mathbf{l})}\mathbf{n}$. The denominator $1 + \varepsilon(\mathbf{n} \cdot \mathbf{l})1 = 1$ when $\mathbf{n} \cdot \mathbf{l} = \mathbf{0}$.

### *Periodic interfaces: satellite positions*

A martensitic laminate stacks two such domains in an alternating sequence: widths $L_1$ and $L_2$ following one another along $\mathbf{n}$, with period $L = L_1 + L_2$. The displacement is now $\mathbf{u} = s(\xi)\mathbf{l}$, where the scalar profile $s(\xi)$ repeats every period[1], $s(\xi + L) = s(\xi)$. Because the displacement phase is periodic in $\xi$, it can be expanded in a Fourier series,

$$e^{i(\mathbf{q}\cdot\mathbf{l})s(\xi)} = \sum_p c_p\,(\mathbf{q}) e^{ipK_0\xi}, \qquad K_0 = \frac{2\pi}{L},$$

with Fourier coefficients

$$c_p(\mathbf{q}) = \frac{1}{L}\int_0^L e^{i(\mathbf{q}\cdot\mathbf{l})s(\xi)}\, e^{-ipK_0\xi} d\xi.$$

Inserting the series into $A(\mathbf{q})$ and using $\xi_j = \mathbf{n} \cdot \mathbf{r}_j^0$,

$$A(\mathbf{q}) = \sum_p c_p\,(\mathbf{q}) \sum_j f_j\, e^{i(\mathbf{q}+pK_0\mathbf{n})\cdot\mathbf{r}_j^0}.$$

The amplitude is now a sum of terms indexed by the integer $p$. Each term is the scattering from the undistorted average lattice, evaluated at the shifted momentum $\mathbf{q} + pK_0\mathbf{n}$ and weighted by $c_p(\mathbf{q})$. The inner sum over atoms produces a sharp reflection wherever this shifted momentum coincides with a parent reciprocal-lattice point, that is, at

$$\mathbf{q} = \mathbf{G} - pK_0\mathbf{n}.$$

These are the satellite peaks: an evenly spaced array of reflections along $\mathbf{n}$, separated by $K_0$ and centred on the parent reflection $\mathbf{G}$. The pinning of this array to $\mathbf{G}$ reflects substrate coherence: the individual domains shear in opposite senses, but their average over a period carries no net strain or macroscopic rotation relative to the parent lattice, which remains coherent with the substrate. The average lattice therefore coincides with the parent, and $\mathbf{G}$ is fixed. The satellite positions are set by the period $L$ alone; the displacement profile does not shift them and enters only through the intensities, $I_p \propto |c_p(\mathbf{q})|^2$. As the transformation proceeds, the satellite positions remain fixed while their intensities evolve with the separation between the two domains' reciprocal lattices.
The total scattered intensity is fixed, whatever the displacement. The modulation $e^{i(\mathbf{q}\cdot\mathbf{l})s(\xi)}$ has unit magnitude, so Parseval's theorem gives

$$\sum_p |\, c_p(\mathbf{q})|^2 = \frac{1}{L}\int_0^L |\, e^{i(\mathbf{q}\cdot\mathbf{l})s(\xi)}|^2 d\xi = 1 \qquad \text{for any } \mathbf{q}\cdot\mathbf{l}.$$

[1] If the profile instead increased by a fixed amount each period, $s(\xi + L) = s(\xi) + C$, subtracting the linear ramp $C\xi/L$ restores periodicity. This ramp is a uniform shear of the entire crystal and displaces every reciprocal-lattice point by a common vector, leaving the satellite pattern unchanged.

The displacement conserves the total scattered intensity and redistributes it between the parent reflection ($p = 0$) and the satellites ($p \neq 0$). The following calculation determines this distribution.

### *Intensity of satellite peaks*

Within each domain the profile is a straight line, of slope $\varepsilon_k$ in domain $k$. Continuity of $s$ at the interfaces, together with periodicity, forces

$$\varepsilon_1 L_1 + \varepsilon_2 L_2 = 0,$$

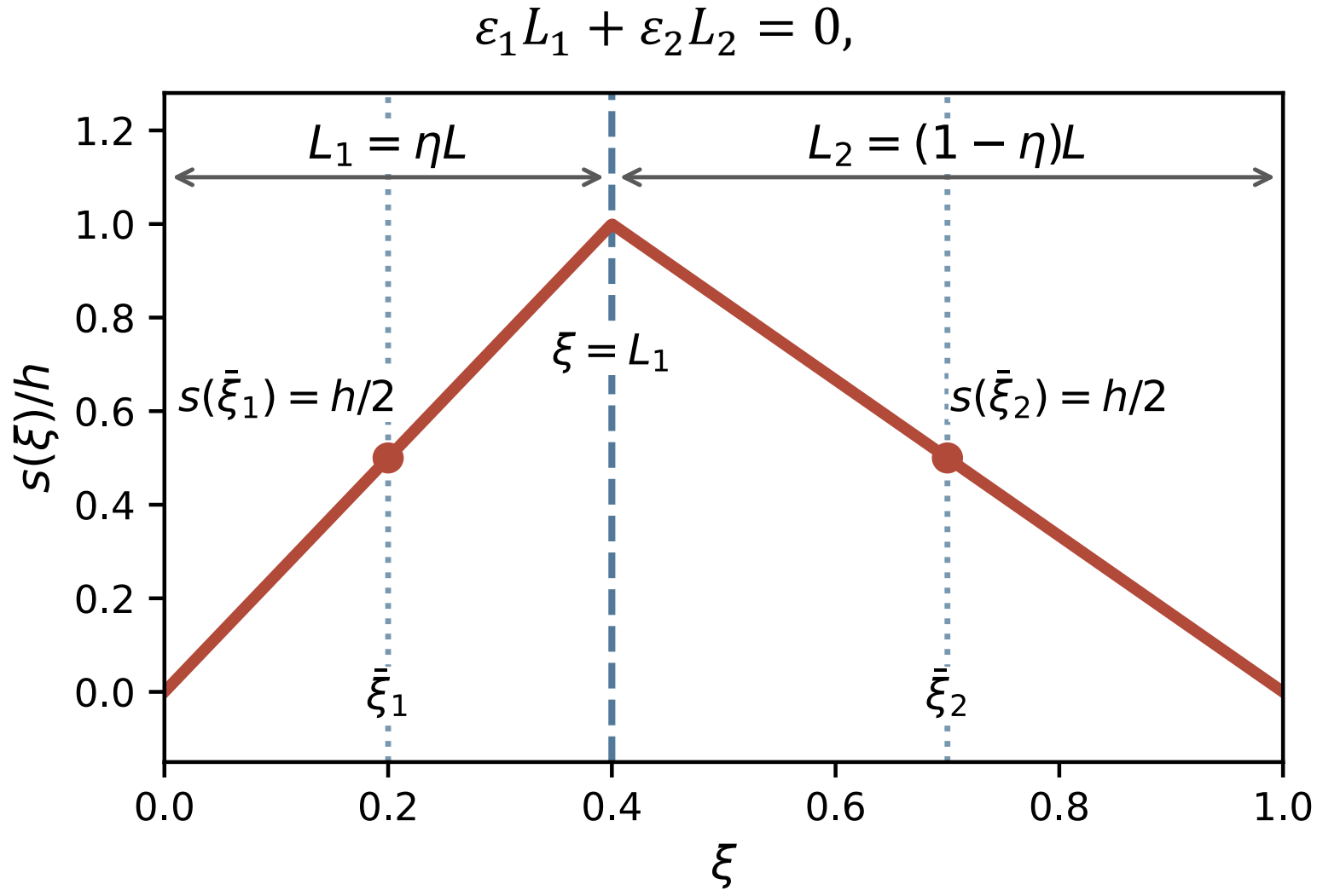

so the two domains shear in opposite senses and $s$ is a triangle wave: it rises across the first domain to a peak displacement $h = \varepsilon_1 L_1 = -\varepsilon_2 L_2$ at the interface, then falls back to zero across the second (see schematic above). The midpoint of each domain lies at

$$\bar{\xi}_1 = L_1/2\,, \qquad \bar{\xi}_2 = L_1 + L_2/2,$$

the center of its own (generally unequal) width. Measuring the displacement phase from each midpoint, the phase in domain $k$ is $(\mathbf{q}\cdot\mathbf{l})s(\xi) = -\Delta_k(\xi - \bar{\xi}_k) + (\mathbf{q}\cdot\mathbf{l})s(\bar{\xi}_k)$, with

$$\Delta_k \equiv \boldsymbol{\Delta}_k \cdot \mathbf{n} = -\varepsilon_k(\mathbf{q}\cdot\mathbf{l}),$$

the same per-domain shift as in the single-interface case. The Fourier integral splits at the interface into one piece per domain, $c_p = c_p^{(1)} + c_p^{(2)}$, with[2]

[2] Each domain retains its own integrated intensity, centered on the domain's reflection $\mathbf{G} + \boldsymbol{\Delta}_k$. The interference between the domains determines only how this fixed intensity is distributed among the satellites without transferring weight between the two reflections.

$$c_p^{(k)}(\mathbf{q}) = \frac{1}{L}\int_{\text{domain } k} e^{-i\Delta_k(\xi-\bar{\xi}_k)}\, e^{-ipK_0\xi} d\xi = \frac{L_k}{L} e^{i\phi_k}\text{sinc}[1/2\,(pK_0+\Delta_k)L_k],$$

where $\text{sinc}(x) = \frac{\sin x}{x}$ and $\phi_k = -pK_0\bar{\xi}_k + (\mathbf{q}\cdot\mathbf{l})s(\bar{\xi}_k)$.

The first term in $\phi_k$ records the location of the domain center along the stacking direction; the second is the phase the displacement contributes at the domain's midpoint. Both midpoints lie at the same height on the triangle, $s(\bar{\xi}_1) = s(\bar{\xi}_2) = h/2$, therefore, the second term is the same for both domains. Defining

$$\chi \equiv (\mathbf{q}\cdot\mathbf{l})s(\bar{\xi}_k) = 1/2\,(\mathbf{q}\cdot\mathbf{l})h,$$

and $\eta = \frac{L_1}{L}$, and using $\Delta_1 L_1 = -2\chi$, $\Delta_2 L_2 = +2\chi$, we rewrite $\phi_1 = -\pi p\eta + \chi$ and $\phi_2 = -\pi p(1+\eta) + \chi$. The phases of the two domains differ by a fixed amount,

$$\phi_2 - \phi_1 = -\pi p,$$

The definition of $\chi$ and $\eta$ allow us to rewrite the arguments of the sinc functions to $u = 1/2\,(pK_0+\Delta_1)L_1 = \pi p\eta - \chi$ and $v = 1/2\,(pK_0+\Delta_2)L_2 = \pi p(1-\eta) + \chi$. The sum of both simplifies to $u + v = \pi p$ for arbitrary $\chi$ and $\eta$. The reason is geometric: the two midpoints are exactly half a period apart, $\bar{\xi}_2 - \bar{\xi}_1 = L/2$, so their position phases differ by $e^{-ipK_0L/2} = e^{-i\pi p}$, while the displacement phase is shared and cancels. Pulling out the common factor $e^{i\phi_1}$ and writing the second domain's center as $v = \pi p - u$, yields

$$c_p = e^{i\phi_1}\left[\eta\frac{\sin u}{u} + (1-\eta)e^{-i\pi p}\frac{\sin(\pi p - u)}{\pi p - u}\right].$$

The half-period phase and the shifted sine combine in one step. Using $\sin(\pi p - u) = \sin(\pi p)cos(u) - \sin(u)cos(\pi p)$ for integer $p$ we rewrite

$$e^{-i\pi p}\sin(\pi p - u) = -\sin u,$$

and

$$c_p = e^{i\phi_1}\sin u\left[\frac{\eta}{u} - \frac{1-\eta}{v}\right] = e^{i\phi_1}\sin u\,\frac{\eta v - (1-\eta)u}{uv}.$$

The bracket simplifies using $u + v = \pi p$,

$$\eta v - (1-\eta)u = \eta(u+v) - u = \pi p\eta - u = \chi,$$

leaving a single closed form, valid for every order $p$ and any $\mathbf{q}\cdot\mathbf{l}$:

$$c_p(\mathbf{q}) = e^{i\phi_1}\frac{\chi\sin u}{uv} = e^{i\phi_1}\frac{\chi\sin(\pi p\eta - \chi)}{(\pi p\eta - \chi)[\pi p(1-\eta) + \chi]},$$

with intensity

$$I_p \propto |c_p|^2 = \frac{\chi^2\sin^2(\pi p\eta - \chi)}{(\pi p\eta - \chi)^2[\pi p(1-\eta) + \chi]^2}.$$

*Coincident reciprocal lattices* ($\mathbf{q}\cdot\mathbf{l} = 0$).

When $\mathbf{q}\cdot\mathbf{l} = 0$ the displacement does not enter the phase ($\chi = 0$), and the two domains' reciprocal lattices coincide with the parent. For every $p \neq 0$ the prefactor $\chi$ vanishes while the denominator remains finite, so $c_p = 0$: the satellites are extinguished by destructive interference between the domains. Only the parent reflection remains at $p = 0$ with $c_0 = 1$.

*Separated reciprocal lattices* ($\mathbf{q} \cdot \mathbf{l} \neq 0$).

A non-zero $\mathbf{q} \cdot \mathbf{l}$ separates the two domains' reciprocal lattices and lifts the cancellation. For small displacement ($\chi \ll 1$) the satellite amplitude is linear in $\chi$: expanding $c_p$, the numerator $\chi \sin(\pi p \eta - \chi) \to \chi \sin(\pi p \eta)$ and the denominator $\to \pi^2 p^2 \eta(1-\eta)$, so

$$c_p \simeq e^{i\phi_1} \frac{\chi \sin(\pi p \eta)}{\pi^2 p^2 \eta(1-\eta)} = i(\mathbf{q} \cdot \mathbf{l}) s_p, \qquad |s_p| = \frac{h|\sin(\pi p \eta)|}{2\pi^2 p^2 \eta(1-\eta)},$$

where $s_p$ is the $p$-th Fourier coefficient of the displacement profile $s(\xi)$. The intensity then separates into two factors,

$$I_p \propto |\mathbf{q} \cdot \mathbf{l}|^2 |s_p|^2 = \frac{|\mathbf{q} \cdot \mathbf{l}|^2 h^2 \sin^2(\pi p \eta)}{4\pi^4 p^4 \eta^2 (1-\eta)^2}.$$

Consistent with the sum rule above, the intensity transferred from the parent reflection appears in the satellites: the parent retains $|c_0|^2 = \mathrm{sinc}^2 \chi$, and the satellites account for the remainder $1 - \mathrm{sinc}^2 \chi$. Increasing $\mathbf{q} \cdot \mathbf{l}$ transfers intensity from the parent reflection to the satellites, each scaling as $\chi^2 \propto |\mathbf{q} \cdot \mathbf{l}|^2$, with successive orders decreasing as $p^{-4}$ and modulated by $\sin^2(\pi p \eta)$.

The leading factor $|\mathbf{q} \cdot \mathbf{l}|^2$ is a selection rule arising from the rank-one nature of the displacement. Because the atomic displacement lies along the single direction **l**, it couples to the scattering only through the projection $\mathbf{q} \cdot \mathbf{l}$. A reflection with $\mathbf{q} \perp \mathbf{l}$ is therefore insensitive to the displacement and exhibits no satellites; a reflection with $\mathbf{q} \parallel \mathbf{l}$ exhibits the maximum satellite intensity. This factor scales all orders uniformly, modulating the entire satellite ladder.

## Description of the Symmetry Quotient

The rank-one selection rule established above makes each twin variant's satellite amplitude depend on $\mathbf{q}$ only through the projection of $\mathbf{q}$ onto that variant's displacement direction $\mathbf{l}$. This holds for every satellite order $p$: the amplitude $c_p \simeq i(\mathbf{q} \cdot \mathbf{l})\, s_p$ carries the orientation dependence entirely in the prefactor $q \cdot l$, common to all $p$, while the shape factor $s_p$ fixes only the relative weights between orders. The two variants, identified by the $012$ and $(0\bar{1}2)$ interfaces (we use orthorhombic notation here and throughout the main text and supplement), carry displacement directions at orientation angles $\theta_{012} = 42°$ and $\theta_{0\bar{1}2} = -42°$ relative to the film normal. Each projection is a cosine of the angle between $\mathbf{q}$ and the corresponding displacement direction (with φ denoting the angle between $\mathbf{q}$ and film normal,

$$\mathbf{q} \cdot \mathbf{l}_{012} = \mid \mathbf{q} \mid \cos(\varphi - \theta_{012}), \qquad \mathbf{q} \cdot \mathbf{l}_{0\bar{1}2} = \mid \mathbf{q} \mid \cos(\varphi - \theta_{0\bar{1}2}).$$

The symmetry quotient, defined as $\mathrm{SQ}_{hkl} = \min(I_1, I_2)/\max(I_1, I_2)$ with $I_1, I_2$ the integrated intensities of the satellite streaks around reflection $hkl$, isolates this angular asymmetry (we take the satellites at $p = 3$, where the overlap with the other diagonal is smallest and signal still sufficient). This yields

$$\mathrm{SQ}_{hkl} = \frac{\min\left(\cos^2(\varphi - \theta_{012}), \cos^2(\varphi - \theta_{0\bar{1}2})\right)}{\max\left(\cos^2(\varphi - \theta_{012}), \cos^2(\varphi - \theta_{0\bar{1}2})\right)}.$$

Figure S2 shows that this approximation works up to a value of $h = 2$.

## Supplementary Figures

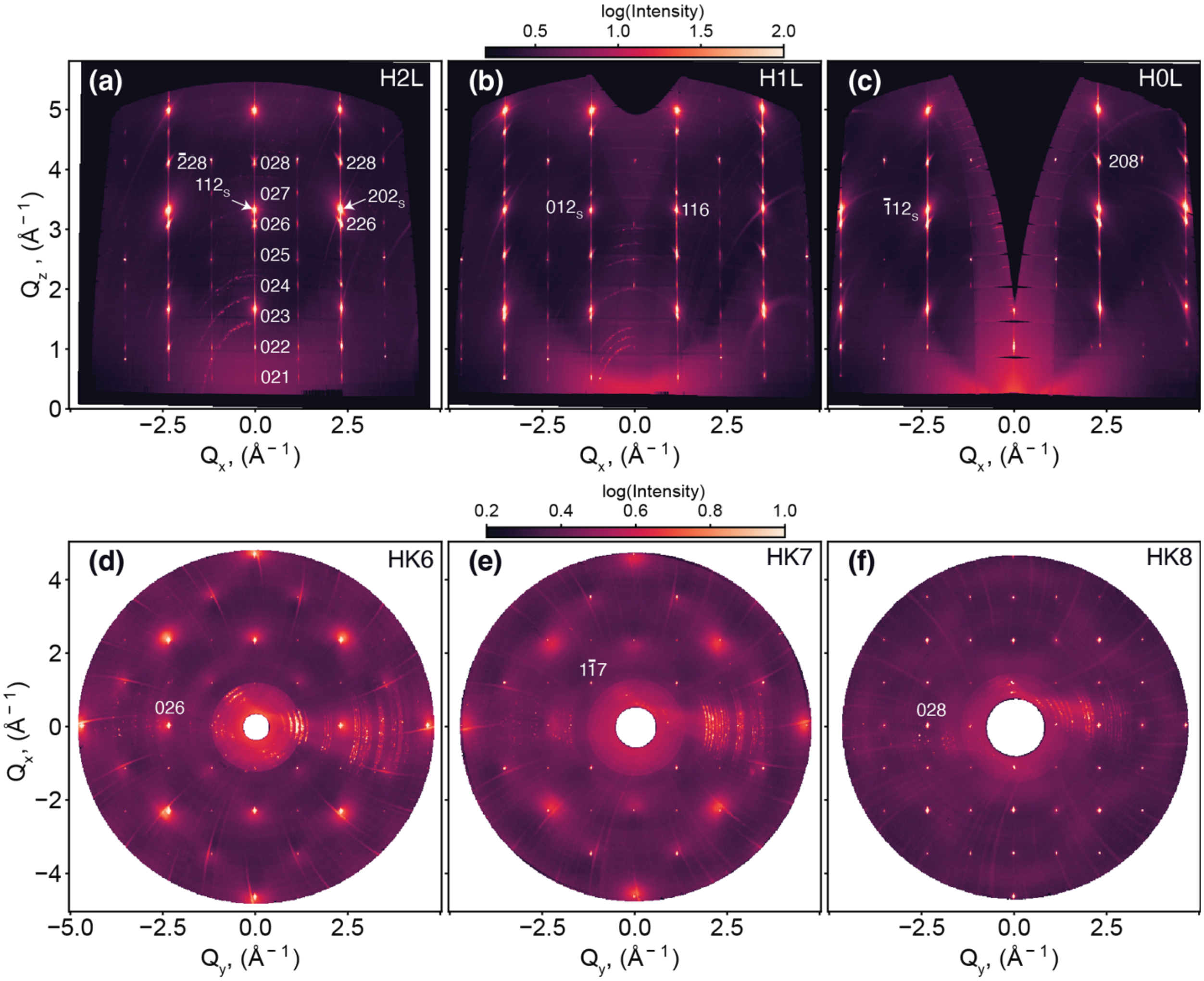

**Figure S1: 2D slices through 3D reciprocal space at various fixed values of $Q_y$ (a-b) and $Q_z$ (d-e).** The Miller indices for selected peaks are indicated. The intensity is shown on a logarithmic scale.

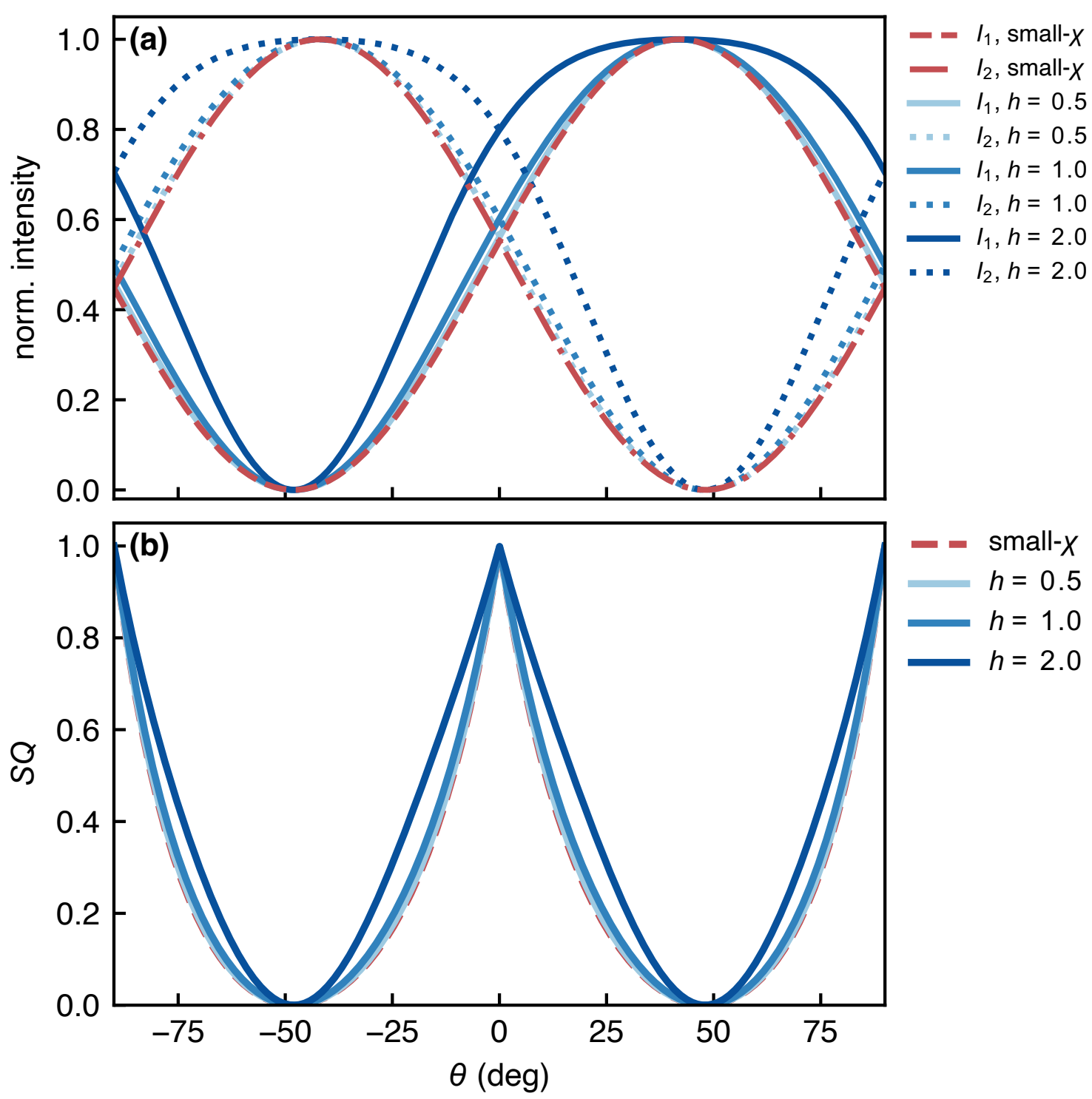

**Figure S2. Satellite intensities and symmetry quotient versus momentum-transfer orientation.** The two twin variants are characterized by displacement directions at $\theta_0 = \pm 42°$. Here, $\theta$ denotes the orientation of $\mathbf{l}$ with respect to the film normal, and $h$ is the maximum magnitude of $\chi = \mathbf{q} \cdot \mathbf{l}$. (**a**) Normalized satellite intensities $I_1$ and $I_2$ for the two variants as a function of $\theta$. Red dashed and dash-dotted curves show the small-$\chi$ approximation, $I \propto (\mathbf{q} \cdot \mathbf{l})^2$, for the two variants. Blue curves show the full expression $I \propto \frac{\chi^2 \sin^2(\pi p \eta - \chi)}{(\pi p \eta - \chi)^2 , [\pi p (1-\eta) + \chi]^2}$ for $h = 0.5$, 1.0, and 2.0. Solid and dotted lines correspond to the two variants. (**b**) Symmetry quotient $Q = I_{\min}/I_{\max}$ calculated from the same intensities. The red dashed curve is the small-$\chi$ approximation (overlayed with $h = 0.5$), while the blue curves are obtained from the full expression for $h = 0.5$, 1.0, and 2.0. The close agreement between the exact and approximate results, particularly for $h \lesssim 1$, demonstrates that the leading-order $(\mathbf{q} \cdot \mathbf{l})^2$ approximation captures the symmetry quotient accurately over the experimentally relevant range. The lattice tilt reported in [9] is $\varepsilon_k = \pm 0.01$ rad, and domain is $L_1 = L_2 \approx 50$ Å (see TEM in Fig. S3). At the highest measured $q \approx 4\text{Å}^{-1}$, 008 peak, we estimate $h \approx 0.01 \times 50\text{Å} \times 4\text{Å}^{-1} = 2$, and all other values of $h$ to be smaller.

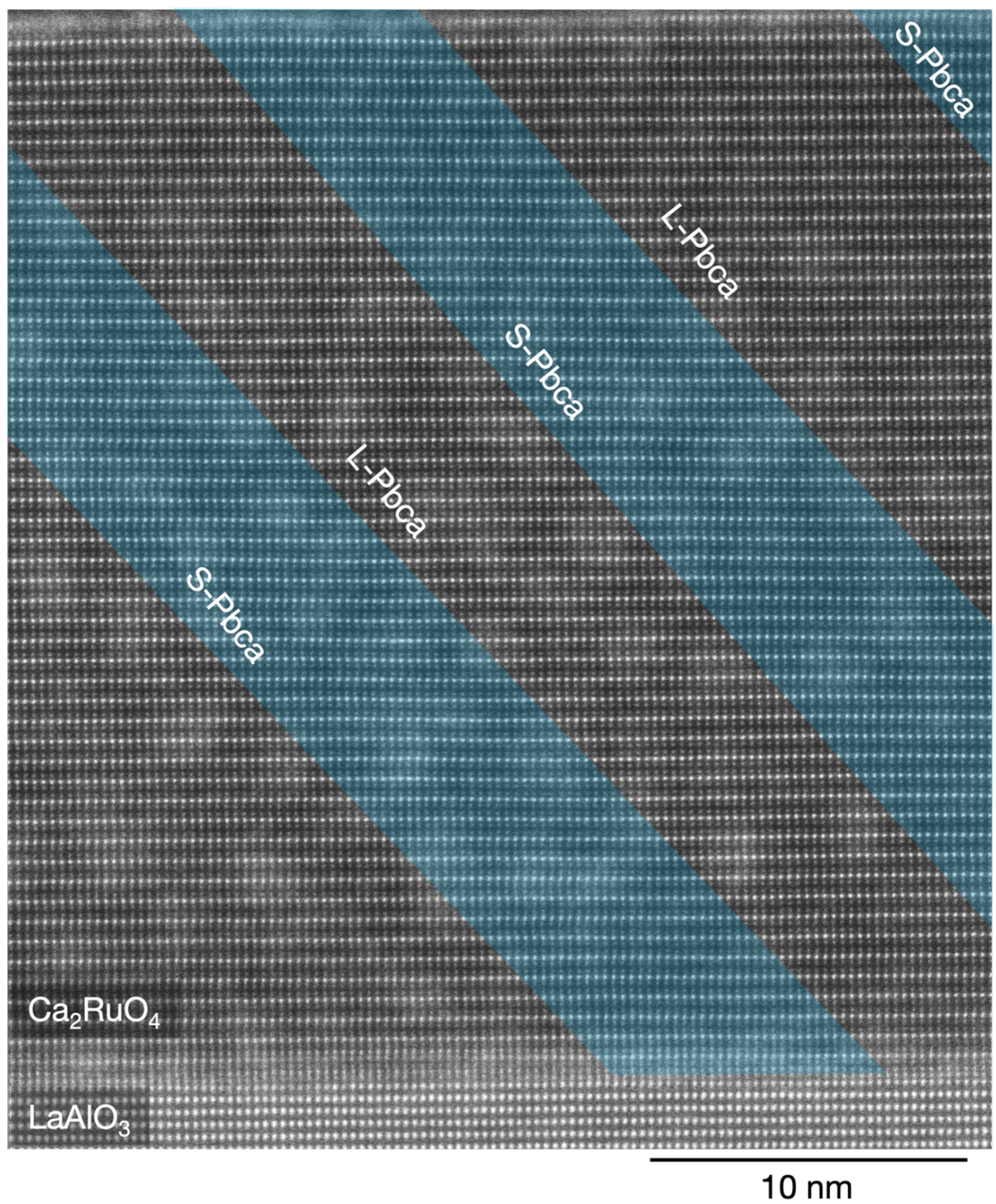

**Figure S3: Scanning transmission electron microscopy** image acquired in projection along the [100] zone axis at 100 K. S (blue overlay) and L-*Pbca* phase regions identified via strain mapping with geometric phase analysis are indicated [9,10]. Atomic planes are continuous across the boundaries between the two phases and no defects are present, identifying the interfaces as coherent in this projection. The data shown is for a film that is thicker than the one studied with X-rays.

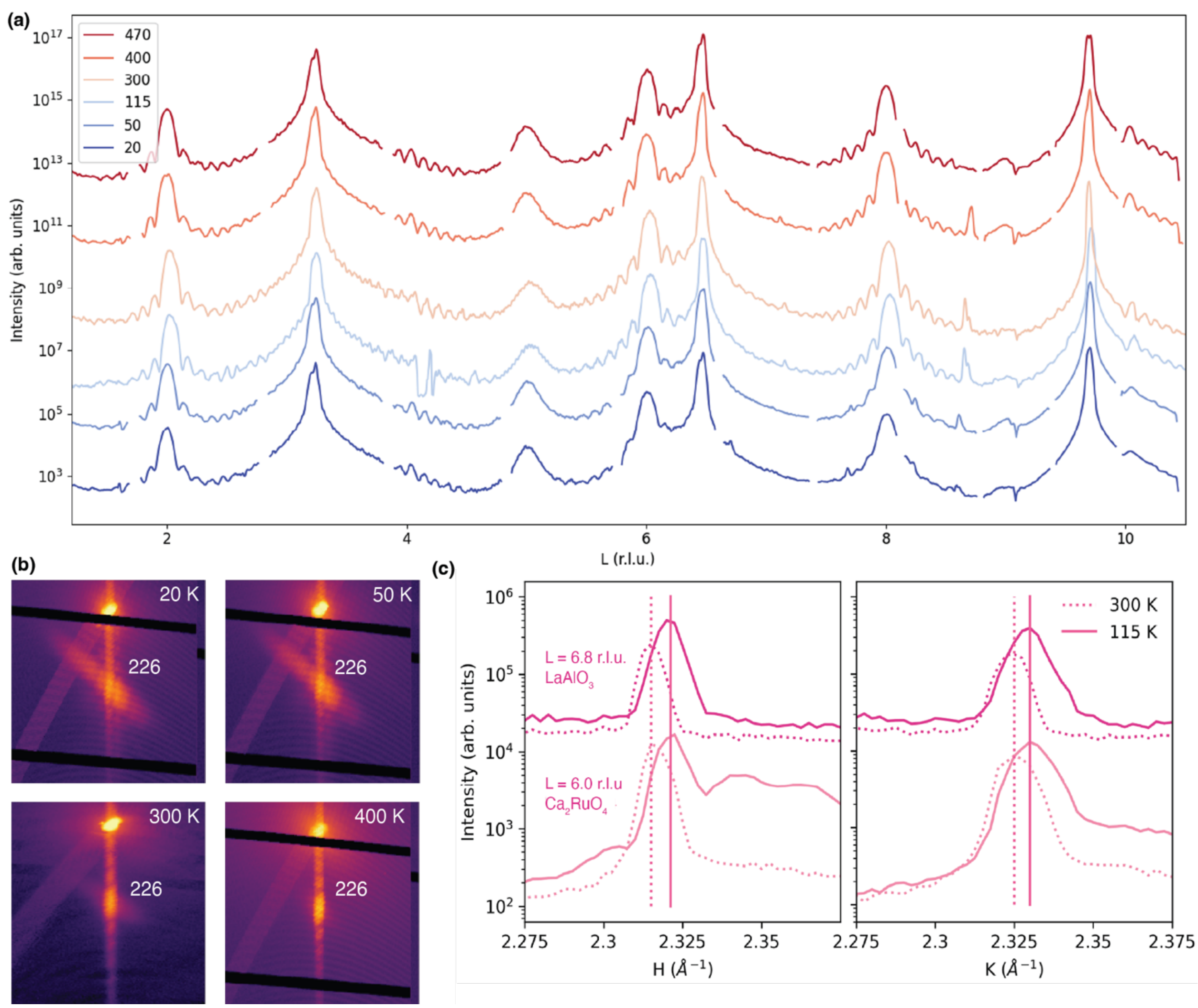

**Figure S4: (a)** Diffraction along the 20L rod measured at different temperatures with no significant intensity changes across all peaks (202 peak is at L=2). The curves are displaced vertically for better visibility. **(b)** The reciprocal space around intensity around the 226 Bragg reflection in the k-l plane for different temperatures. The satellite intensity disappears at higher temperatures. **(c)** The position of the $LaAlO_3$ substrate crystal truncation rod (CTR) running parallel to [001], and the 226 Bragg peak in H and K. The substrate CTR and the film peak overlap in the L-fixed plane for both temperatures, 300 and 115 K, confirming that the films are commensurately strained (the in-plane lattice constants of the film determined by the position in H and K, match the lattice constants of the substrate).

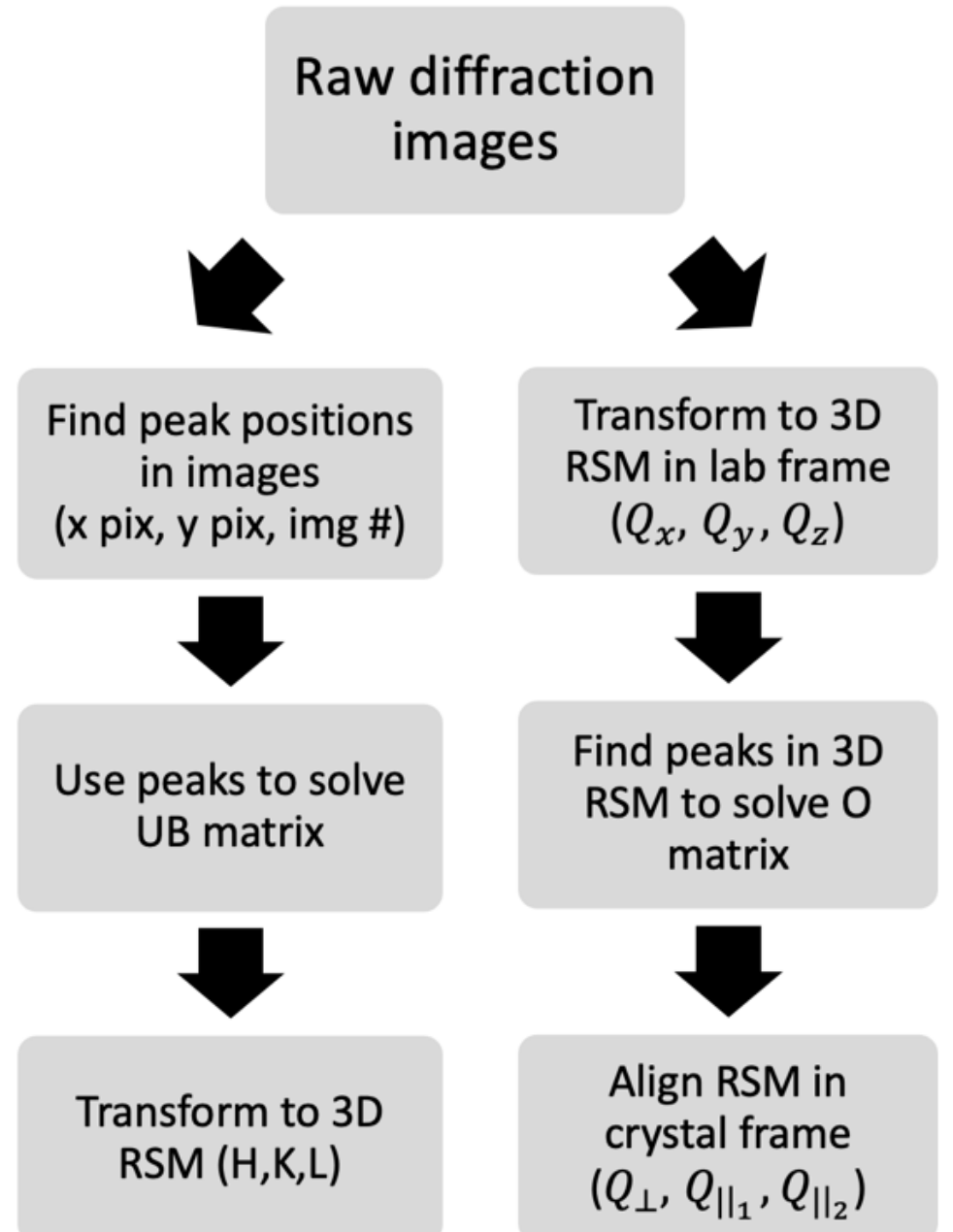

**Figure S5:** A flow chart for the two algorithms transforming raw data to 3D reciprocal space volume.

**References**

[1] O. Friedt, M. Braden, G. André, P. Adelmann, S. Nakatsuji, and Y. Maeno, Structural and magnetic aspects of the metal-insulator transition in Ca2−xSrxRuO4, Physical Review B **63**, 174432 (2001).

[2] W. R. Busing and H. A. Levy, Angle calculations for 3- and 4-circle X-ray and neutron diffractometers, Acta Cryst **22**, 457 (1967).

[3] J. Kieffer and D. Karkoulis, PyFAI, a versatile library for azimuthal regrouping, J. Phys.: Conf. Ser. **425**, 202012 (2013).

[4] P. Virtanen et al., SciPy 1.0: fundamental algorithms for scientific computing in Python, Nat Methods **17**, 261 (2020).

[5] G. Winter et al., *DIALS* : implementation and evaluation of a new integration package, Acta Crystallogr D Struct Biol **74**, 85 (2018).

[6] A. Khachaturyan, *Theory of Structural Transformations in Solids*, 1st ed. (Dover Publications, 2008).

[7] Y. U. Wang, Diffraction theory of nanotwin superlattices with low symmetry phase, Phys. Rev. B **74**, 104109 (2006).

[8] B. E. Warren, *X-Ray Diffraction* (Dover Publications, New York, 1990).

[9] Z. Shao et al., Real-space imaging of periodic nanotextures in thin films via phasing of diffraction data, Proc Natl Acad Sci U S A **120**, e2303312120 (2023).

[10] N. Schnitzer, G. Powers, B. H. Goodge, E. Bianco, I. E. Baggari, and L. F. Kourkoutis, Atomic-Resolution Imaging of Phase Transitions in Strongly Correlated Oxides with Continuously Variable Temperature Cryo-STEM, Microscopy and Microanalysis **29**, 1688 (2023).